\documentclass[preprint2]{aastex}

\slugcomment{To appear in Astronomical Journal}

\shorttitle{Star Formation in Satellite Galaxies}

\shortauthors{Guti\'errez et al.}

\begin{document}

\title{Star Formation in Satellite Galaxies\altaffilmark{1}}

\author{C. M. Guti\'errez}
\affil{Instituto de Astrof\'\i sica de Canarias, E-38205 La Laguna, Tenerife, Spain}
\email{cgc@iac.es}

\author{M. S.  Alonso\altaffilmark{2}}
\affil{Consejo Nacional de Investigaci\'on Cient\'\i ficas y T\'ecnicas, Argentina}
\email{salonso@casleo.gov.ar}

\author{J. G. Funes, SJ}
\affil{Vatican Observatory, Univ. of Arizona, Tucson, AZ 85721, USA}

\and

\author{M. B. Ribeiro}
\affil{Instituto de F\'\i sica, Universidade Federal do Rio de Janeiro, Brazil}

\altaffiltext{1}{Based on observations with
the VATT: the Alice P. Lennon Telescope and the Thomas J. Bannan Astrophysics
Facility.}
\altaffiltext{2}{Marie Curie Fellowship, Instituto de Astrof\'\i sica de Canarias}

\begin{abstract}
We present narrow-band observations of the H$\alpha$ emission in a sample of 31 satellite
 orbiting isolated giant spiral  galaxies. The sample studied spans the range
$-19<M_B <-15$ mag. The H$\alpha$ emission was detected in all the spiral and irregular
objects with fluxes in the range $1.15-49.80\times 10^{-14}$ erg cm$^{-2}$ s$^{-1}$. 
The average and maximum values for the current star formation  rates are 0.68 and   3.66
M$_\sun$ yr$^{-1}$ respectively. Maps of the spatial distribution of ionized gas  are
presented. The star-forming regions show a rich structure in which frequently discrete
complexes are imposed over more diffuse structures. In general, the current star
formation rates are smaller that the mean values in the past obtained from the current
stellar content; this probably indicates a declining rhythm with time in the generation
of new stars. However, the  reserve of gas  is enough to continue fueling the current
levels of  star formation activity for at least another Hubble time. Four of the objects
(NGC~2718b, NGC~4541e, NGC~5965a$_1$ and NGC~5965a$_2$) with higher current star formation
rates show
clear signs of interaction with close companions of comparable brightness at projected
distances of 25, 20 and 2 kpc respectively. The only two galaxies in our sample that do
not show star formation activity are members of these interacting systems, and it is
unclear if this is a consequence of intrinsic properties (both are Hubble early types) or
if it is related with possible disruption of the external parts due to the interaction.
In the case of the pair NGC~2718a-b there are indications of gas transport between both
galaxies. 

\end{abstract}

\keywords{galaxies: fundamental parameters -- galaxies: photometry -- galaxies:
structure -- galaxies: star formation rate}

\newpage

\section{Introduction}
Cold dark matter cosmologies predict a hierarchical scenario (e.g., White \& Rees
1978) in which small structures are formed first, and then by several processes of
merging, accretion, etc., larger structures are generated. Currently, one of the most
discussed questions in this field is the so-called ``missing satellite problem''; in fact
semi-analytic models (Kauffmann et al.\ 1993) and numerical simulations (Klypin et al.\
1999; Moore et al. 1999) of  small structure formation in cold dark matter cosmologies 
predict a number of satellites in the halos of large galaxies an order of magnitude
larger than the observed counts in the Local Group. A possible observational
incompleteness of factor up to three of the known population of dwarfs orbiting the
Milky Way,  as has been evaluated  by Willman et al.\ (2004), is  not  enough to solve the
discrepancy.  It has also been proposed that star formation in low-mass structures could
be inhibited by a strong photo-ionizing  background (Somerville 2002),  especially at low
redshift (Dijkstra et al.\ 2004).  

Reconstructing the evolutionary history of satellites in  galactic halos are, then,
essential for validating the main predictions of these models. This requires  study of the
role of physical processes such as interactions, mass losses, morphological transformation 
and star formation. Obviously, the best studied are the Milky Way and Andromeda halos,
although extending the study to other host galaxies is necessary for studying
possible variations  from halo to halo, which probably  depends on the host mass, the
merging history, and the environment.  That was our motivation to start an observational
program that comprises photometry in optical broad- and narrow-band filters for a sample
of satellites orbiting external giant spiral galaxies. In previous articles (Guti\'errez
et
al.\ 2002;  Guti\'errez \& Azzaro 2004) we have analyzed the morphology and photometry  of
about 60  such objects.  This analysis enabled us to validate and extend the relations
found in the  satellites of the Local Group. In this article, we use narrow-band
observations in H$\alpha$ to estimate  the current star formation rate of a subsample
comprising most of the late type objects. We analyze the evolution of  star-forming
activity with time, and the relation with  morphological type, HI mass, and
environment.  The paper is organized as follows. Section~2 presents the
observations  and data reduction; Section~3 and Appendix~A present the images and 
analyze qualitatively the H$\alpha$ maps of each object; the star formation rate estimates 
are presented in Section~4; the relation
between interactions and star formation,  and the cosmological evolution of  star
formation  activity are discussed in  Sections~5 and 6; finally Section~7 presents the
conclusions.

\section{Observing program}

\subsection{The Sample} 
The initial sample is the catalogue of satellite galaxies compiled by  Zaritsky et al.\
(1997). The catalogue contains 115 objects orbiting  69 primary isolated spiral
galaxies. Basically, the satellites were selected according to their relative
brightness (at least  2.2 magnitudes fainter than their parents), projected distances
($<500$ kpc) and relative velocity ($<500$ km s$^{-1}$)  from the primaries. The objects
analyzed here correspond to a subsample comprising  most of the objects classified by
Guti\'errez \& Azzaro (2004) as spirals or irregulars. In total, 31 objects have been
observed and analyzed (most of 
the late type objects  observable from the northern hemisphere). The
objects span a large range in luminosity ($-19 < M_B < -15$), and constitute a 
sample useful for statistical studies of the population of late-type galaxies present in
the halos of large spiral galaxies. 

\subsection{Observations and data reduction}
\begin{figure}
\plotone{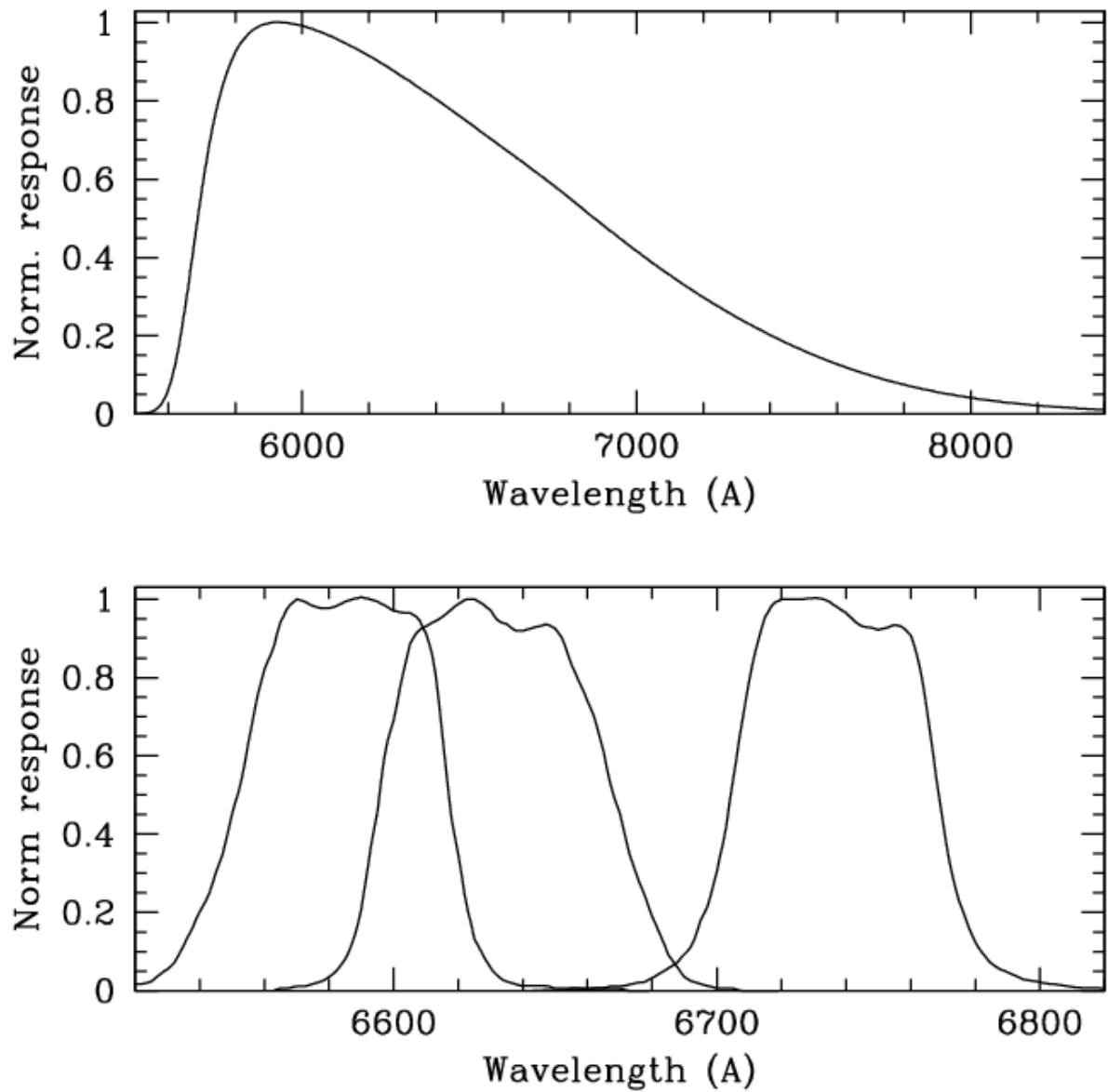}
\caption{Normalized response of the R broad band filter ($top$), and the three narrow
band filters ($bottom$) used during the observations presented in this paper.}
\label{filtros}
\end{figure}

The H$\alpha$ images and the broad-band for continuum estimation  were acquired during
three observing runs seven nights each in December 2001, May 2002, and December 2002 with
the 1.8 m Vatican Advanced Technology Telescope (VATT) at the Mt Graham International
Observatory. Most of the observations analyzed here were obtained in the first two runs
and
in photometric conditions (for only for two of the galaxies  was it possible to make an
absolute direct calibration, see next section). A back-illuminated 2048 $\times$ 2048
Loral CCD
was used as the detector at the aplanatic Gregorian focus, $f$/9. It yielded a field of
view
of 6.4$'$ x 6.4$'$ with an image scale of 0.4 pixel$^{-1}$ after 2 $\times$ 2 pixel
binning. The
seeing varied between 1.1$''$  and 2.3$''$ with a mean value of 1.5$''$. For each galaxy
we
have obtained typically 3 $\times$ 1800 s narrow-band images using an appropriate
interference
filters with $\sim 70$ \AA~widths that  isolate the  spectral region characterized by the
redshifted H$\alpha$ and [N~{\scriptsize II}]$\, \lambda 6548, 6583$ \AA\ emission lines.
To cover the range in velocity spanned for the galaxies presented in this work, three
filters were needed with central wavelength of  6584.7, 6632.8 and 6736.2 \AA~(another
filter exists centered at 6683.1 \AA\ was not necessary for any of the galaxies analyzed
in
this work). The nominal normalized spectral response of these filters are presented in
Figure~\ref{filtros}. Table~1 presents a summary of the observations and properties of the
images obtained. Each filter is denoted according to its central wavelength in nm.  

For absolute astronomical calibration we observed spectrophotometric standard stars from
the list provided by Oke (1990). In general, we observed one of these stars just before or
after the narrow-band observations for each galaxy. In each case, stars were selected with
similar 
airmass rather than to calibrate the target. The exposure times for these stars
range between 12 and 300 s, depending obviously on the  magnitude of the star. These
details are shown in Table~1. The data were reduced using  IRAF packages\footnote{IRAF is
the Image Reduction and Analysis Facility, written and supported by the IRAF programming
group at the National Optical Astronomy Observatory (NOAO) in Tucson, Arizona.}. We
performed standard data reduction, comprising, bias subtraction, flat-field
correction using sky twilight observations in the appropriate filters, alignment of $R$
and
narrow-band images, and co-addition of  the  narrow-band images. This combination
eliminates most of the artifacts due to bad pixels or cosmic rays. 

To isolate the possible H$\alpha$ emission of the targeted galaxies, the contribution of
the stellar continuum emission of the galaxy needs to be removed. This was done by
appropriately scaling
the $R$-band image and subtracting it from the H$\alpha$ images. The scaling
factor was estimated by comparing the brightness of several field stars in the broad and
in
the narrow bands respectively. We measured the flux of these stars in a circular aperture
$\sim$6 FWHM of the image. Depending on the field, the number of stars considered was in
the
range 3--12, with a typical number of 5. 

\section{Images of the satellite galaxies}
The $R$-band and H$\alpha$ images, after continuum subtraction, are presented in
Figures~2, 3 and 4. We have detected H$\alpha$ emission in all but two galaxies. For most
of
the galaxies, the H$\alpha$ emission  shows a complex morphology in which it is possible
to
discern diffuse and discrete structures. A description of these images is presented in
Appendix~A. The identification of interacting systems is based  on spatial
proximity and on the presence of features in the $R$ band, such as a bridge in the case of
the
pair NGC~2718a-b, tidal streams in the pair NGC~4541b-e, and distortion of
the outer isophotes in the case of the two components of NGC~5965a. For the other galaxies
there is also some evidence of current or past interactions on the basis of the proximity
to the parent (NGC~3154a), possible companions with unknown redshift (NGC~3735a, 
NGC~4030b,
NGC~4541a and NGC~6181a), knots differentiated from the main body of the galaxy having
strong star formation activity (NGC~1961b,  NGC~5899a and NGC~7678a), or the presence
of possible tidal
galaxies (as in NGC~2775c). 

\section{H$\alpha$ luminosities and star formation rates}

We use the Sextractor software (Bertin \& Arnouts 1996) to carry out the photometry of the
galaxies in the broad and narrow-band filters. We consider two possible estimates for the
total flux of the galaxies. The first is the integrated flux in a predefined aperture (the
same as for the narrow and broad filter) and the second the flux in the automatic aperture
computed in Sextractor according to the extrapolation  proposed by  Kron (1980). In
general, we note that the first method gives more robust values for those galaxies  that
are more irregular,  while the second is more appropriate
for objects with a well defined regular profile.  In the cases in which a fixed aperture
was chosen, we checked that the estimated fluxes are similar to those computed with the
subroutine PHOT of IRAF. For the absolute calibration and estimation of H$\alpha$
emission,
we follow the procedure detailed in Gil de Paz et al.\ (2003). Because of the width of our
narrow filters, the emission includes also the contribution of [NII]  for which an
accurate
subtraction would require spectroscopic observations.

One of the objects (NGC~4030b) was observed at two epochs (Dec 2001  and May 2002).  We
think that the $\sim$10\% difference between the estimates of H$\alpha$ fluxes in  both
observations  (1.33 and $1.46\times 10^{-14}$ erg cm$^{-2}$ s$^{-1}$ for the observations
of December and May respectively) are representative of the statistical  uncertainty of
our
procedure. Another test was to observe the apparently non-interacting early-type
galaxy NGC~1620b for which we do not  in principle expect significant levels of
star-forming
activity. For this galaxy we estimate an H$\alpha$ flux after continuum subtraction of
$-0.87\times 10^{-14}$  erg cm$^{-2}$ s$^{-1}$, which in absolute value is well below any
of the detections of H$\alpha$ emission. We think that this negative value is not
entirely due to the uncertainties in the estimate, but  indicates
H$\alpha$ in absorption, as is commonly found in early-type galaxies (Kennicutt 1992).

The H$\alpha$ equivalent widths (EWs) represent the ratio between the H$\alpha$
emission and the continuum averaged over the full galaxy.  Figure~\ref{ewtipo} shows
the relation between the relative flux of the H$\alpha$ emission and the
morphological type. The mean values and 1$\sigma$ dispersions in H$\alpha$ EWs are
$16.4\pm 5.1$, $24.3\pm 9.6$, $18.9\pm 7.7$, and  for  types  2 (Irr), 3 (Sb/Sc), and 4
(Sa) respectively.  The mean values found for the different
morphological types of non-interacting objects are  similar,  and then we do not
notice differences in H$\alpha$ EWs with  morphological types a found by
other authors (e.g.\ Kennicutt 1998).  The much  higher  H$\alpha$ EWs values found for
four
of the satellites in interaction  indicate that their activity is  directly related
with current  interaction process.  

The H$\alpha$ EWs were converted in fluxes ($f_{\rm H\alpha}$) following the calibration
explained in Section~2; $f_{\rm H\alpha}$  was then converted into absolute luminosity
according to $L_{\rm H\alpha}=4\pi D^2f_{\rm H\alpha}$. To estimate distances, we took the
recessional velocity of the respective parent galaxy (the  satellite is more affected by
the peculiar velocities).  In any case, the difference in the estimated luminosity would
be
 relevant only for a few  systems at very low redshift and with comparatively large
differences in velocity between the parent and the satellite (NGC~4030b is the extreme
case). For the Hubble constant we assumed  $H_0=72$ km s$^{-1}$ Mpc$^{-1}$. The  H$\alpha$
fluxes and luminosities are presented in Table~2. For the galaxies with H$\alpha$
emission,
the luminosities are in the range $\log L_{\rm H\alpha}$ = 38.85-41.66 erg s$^{-1}$. The
histogram with the distribution of H$\alpha$ luminosities is presented in
Figure~\ref{lumha}.

To convert $L_{\rm H\alpha}$ to current star formation rates (SFRs), we use the
calibration provided
by Kennicutt (1998), SFR ($M_\sun~yr^{-1}$) = 7.9$\times 10^{-42}L_{\rm H\alpha}$ erg
s$^{-1}$,
which assumes a Salpeter (Salpeter 1955) initial mass function between $10^{-1}-10^2$
$M_\sun$. Because we do not have observations of other Balmer lines (specifically
H$\beta$)
we did not  apply any correction for internal extinction. The values obtained for
the current SFR  are presented in Table~3 column 2). Figure~\ref{sfrb} shows SFR
as a function of luminosity in the $B$ band. A significant correlation (although with a
large scatter)  exists. Similar tendency was found in the sample of blue compact
galaxies studied by Kong (2004), and with  the much large sample  of
$\approx$ $10^5$ SDSS galaxies studied by Brinchmann et al.\ (2004). Four of the 
interacting satellite galaxies (indicated by filled circles in the figure) show a  
similar
tendency, although they clearly have higher levels of star formation than non-interacting
galaxies with similar luminosities.

\subsection{Comparison with previous star formation rates estimations}
SFRs of some of the objects analyzed in this work have been previously estimated 
by other
authors. In some cases, their SFRs have been estimated from
H$\alpha$ measurements in narrow-band images or spectroscopy, while in others,  SRFs
have been obtained from  observations in the far-IR or radio. Below we compare our
results with these previous studies compiled from the literature.

$\bullet$ NGC~2718a and NGC~2718b: Mendez et al.\ (1999) have estimated a luminosity of
40.87 erg s$^{-1}$ for the component NGC~2718b, which is in good agreement with our value
of 40.91 erg s$^{-1}$. However, they derived an  EW of 260 \AA~for the H$\alpha$ emission
of
this galaxy, which is a factor $> 2$ larger than our value. Our estimate of H$\alpha$ EW
for this galaxy is also $\sim$2.5 smaller than that presented by Kong et al.\ (2002)
from long-slit observations. However, our EW and H$\alpha$ luminosities  are in very good
agreement with the estimates by Gil de Paz et al.\ (2003), 120 \AA~and  40.85 erg s$^{-1}$
respectively. So, considering only the three measurements based on narrow-band imaging, it
seems that the three estimates of the H$\alpha$ luminosities are in agreement, but they
are in clear disagreement with EWs derived from long-slit spectra. Although the reason for
this discrepancy between the  estimates obtained with each technique is unclear, it could
be related to the  irregular spatial distribution of the H$\alpha$ features. This renders
inaccurate  the extrapolation of the H$\alpha$ intensity from the area covered by the
long-slit of the whole galaxy.

$\bullet$ NGC~5965a$_1$-a$_2$: The estimates carried out previously by other  authors
enclose the H$\alpha$ flux for the whole system. For this, our estimate of the
H$\alpha$  flux is $27.2\times 10^{-14}$ erg s$^{-1}$ cm$^{-2}$ , which is in good
agreement with  the value of 30$\pm3 \times 10^{-14}$ erg s$^{-1}$ cm$^{-2}$ measured
by Gil de  Paz et al.\ (2003). For this pair, 
Bell (2003) obtained an SFR of 0.8 M$_\sun$ yr$^{-1}$ from radio
measurement, while Hopkins et al.\ (2002) obtained  0.8 and 1.7 M$_\sun$ yr$^{-1}$ from
observations  at 60 $\mu$m and 1.4 GHz respectively. However, these authors use a
different distance to the galaxy (53.2 Mpc) instead of the value 47.6 Mpc that we have
used. 
Converting our results to their  distance scale, we would obtain SFR = 0.72 M$_\sun$
yr$^{-1}$,
which is within 25\% of the value derived by these authors from the luminosity at
60 $\mu$m. Given the different uncertainties and assumptions in each of the methods,
we consider both estimates to be in very good agreement. 

\section{Interactions and star formation}
Observations and simulations have  shown that galaxy interactions and mergers are powerful
mechanisms for triggering  star formation (e.g.\ Kennicutt 1998; Donzelli \& Pastoriza
1997;
Barton et al.\ 2000; see, however, Bergvall et al.\ 2003). Lambas et al.\ (2003) and
Alonso et
al.\ (2004) have carried out statistical analyses of star formation in the sample of
galaxies of the 2dF survey (Colless et al.\ 2001),  finding that objects with close
neighbors tend to have a higher star formation activity than those that are isolated. 
This tendency
is more efficient for pairs situated in low density regions. According to the studies by
Sekiguchi \& Wolstencroft (1992) and  Donzelli \& Pastoriza (1997),  enhancement of the
star formation activity is more likely to take  place in both galaxies of the pair but
tends to be higher in the fainter component. However, the
analysis  by  Lambas et al.\ (2003) indicate that a large number of interacting galaxies
do
not show enhanced star formation activity. 

Two basic types of interactions can be differentiated in the halos of giant galaxies;
those between parent and  satellites, and those between satellites. Both in principle can
produce morphological changes and mass losses,  and affect the star formation activity in
the interacting objects; for instance Knebe et al.\ (2005) by using $N$-body simulations
have estimated
that interactions between satellites can account for $\sim$30\% of the total lost masses
through their hectic existence. 

Guti\'errez et al.\ (2002) pointed out two clear cases of interaction (NGC~2718a-b,
NGC~4541b-e)  between pairs of satellites. An additional pair (the two components of
NGC~5965a) was also considered by Guti\'errez \& Azzaro (2004). A relatively high fraction
of other satellites in our sample  show morphologcal signs that could be attributed to
present or  recent interaction (see  Section~3 and Appendix~A), but the possible
companions
have not been accurately identified. Our results confirm that   interactions between
satellites significantly affect   the star formation activity.

According to Guti\'errez \& Azzaro (2004) the difference in magnitudes between the
two members of the pairs NGC~2718ab, NGC~4541be, and NGC~5965a$_1$a$_2$  are $\delta
m\sim 1$ mag in the $R$ band. In the first two cases, a high level of H$\alpha$
emission is detected in just one  member of the interacting system, NGC~2718b  and
NGC~4541e, which are brighter components of their respective pairs.   The other two
galaxies in these pairs, NGC~2718a and NGC~4541b are the only two members of the
sample  analyzed that do not show H$\alpha$ emission. These two galaxies are also
the only objects in our sample that were classified as early types and have been
 included only as members of the interacting pairs. The filaments and bridges seen in
the pairs  NGC~2718a-b and NGC~4541b-e do not show H$\alpha$ emission either.  We
propose that in the case of NGC~2718a the presence of the bridge connecting this
galaxy with the companion, and the comparative large amount of gas (see next
section) in NGC~2718b are signs of mass transfer from one galaxy to the other. This
is probably inhibiting the galaxy formation in the donor  and enhancing it in the
accreting  galaxy. The stripping suffered by NGC~2718a could  also be  responsible for
its morphological evolution toward  early-type objects.  The situation is more
confuse for the pair NGC4541~b-e, where only one of  the two plumes emerging from
NGC~4541b seems to point  towards NGC~4541e and there are no  HI
measurements in the literature for any of the two galaxies of the pair. 

In the pair NGC~5965a$_1$-a$_2$ both members show a high level of star formation,
the brighter galaxy (a$_1$) also being the one that has experienced the more
intense SFR activity. In this
case, the small projected spatial distance, the absence of tidal streams, or bridges and 
the elongation of the isophotes through the respective companion  seem indicate that these
two galaxies are experiencing the first approximation and the gas is being strongly
compressed.\footnote{We consider  the  two components of NGC~5965a to have velocities of
3357 and 3429 km s$^{-1}$ for NGC~5965a$_2$ (SBS~1533+574A) and NGC~5965a$_1$ 
(SBS~1533+574B) respectively (Izotov, Y. 2005 priv.
comm.). Apparently the recessional  velocity quoted in the NED database for the component
SBS~1533+574B is wrong.} We have also computed  the H$\alpha$ emission in an aperture
enclosing both galaxies and  obtained SFR = 0.576 M$_\sun$ yr$^{-1}$. This  is $\sim$5\%
higher than the result obtained by just adding  the H$\alpha$ emission estimated for each
component.  Although compatible with the uncertainties of the method, this excess  could
also indicate the  presence of some H$\alpha$ residual emission  in the  intergalactic
media. The fact that in each of the interacting pairs the SFR activity is more
intense in the brighter component is in disagreement with the results of Donzelli \&
Pastoriza (1997).

Our sample does not contain clear cases of interactions between satellites and
parents, so that it is not possible to estimate the relevance of such interactions to 
star
formation activity. We note, however, that  NGC~3154a,  which is the object in our
sample that has the smaller projected distance from its progenitor (19 kpc), is one of
the more active satellites forming stars. However, we do not notice special signs of
morphological distortion in this satellite.

We can roughly  check  if the abundance of interacting pairs found is reasonable according
to the expectations of numerical simulations in
CDM models. Knebe et al.\ (2004) have estimated that the mean number of satellites that
have
at least an  encounter per orbit with other satellites depends on halo age, and ranges
from
$\sim $4\% (for old halos) to $\sim$58\% (for young halos). These encounters tend to be
more frequent and faster in the internal parts of  halos.  The three interacting pairs
found in a sample of $\sim$60 satellites studied by  Guti\'errez \& Azzaro (2004)  would 
then require very low  velocity encounters, with interactions lasting $\sim$1/2--1/8 of
the orbital
period. This seemsto  contradict the observed differences in radial velocities 
($\sim$50--90 kms$^{-1}$), and projected distances ($\le 20$ kpc) betweeen both
members of each pair. However, larger statistical samples are needed to make a more robust
assesment. 

\section{Current vs. past star formation rates}

The star formation history for satellites in CDM halos have been considered by Mayer
et al.\  (2001). These author show that high and low surface brightness  and
galaxies  (HSB and LSB respectively) react  differently to the interaction with the
central galaxy.
In LSB satellites bursts of activity occur after each pericentric passage, while in
HSB satellites the first burst consumes most of the gas. In both cases the activity is
very inhomogeneus over cosmological times. Following the method proposed by van Zee (2001)
it is possible to estimate the
average star formation rate over the Hubble time from the total mass of stars formed.
The resulting
$<SFR>_{\rm past}$ for the galaxies in our sample are presented in Table~3 column 3, and
are in the range  $\sim$0.04--1.04
M$_\sun$ yr$^{-1}$. For most of the galaxies, these values are larger than the
current star formation rates as estimated from the $H\alpha$ luminosities. The
normalized distributions of the current and past SFR are shown in
Figure~\ref{histsfr}. It is worth noting  that the current SFR spans a large range of
values  extending the distribution to very low values. The histogram of
$<SFR>_{\rm past}$  looks sharper and shifted to higher values. The ratio between
present and average past star formation rates is usually called the birthrate parameter
$b=SFR_{\rm current}/<SFR>_{\rm past}$. The mean value of this parameter is largely
conditioned
by the high values of the starbursting interacting systems, so we think as more
representative of the general distribution, the median, which results $0.68$.

The absolute magnitude of each satellite vs.\ $\log b$ is shown in Figure~\ref{bscalo}.
The
region with higher values of the $b$ parameter is occupied by the four starbursting
interacting galaxies and by NGC~3154a. The $b$ values for these galaxies are in the
range 3--9. We notice a weak trend in the sense that brighter satellites seem to have a
comparatively higher current activity; in fact, four of the five brightest galaxies 
(excluding the interacting systems) have $b\sim 1$, which seems to indicate that these
galaxies are still able to maintain now their mean past star formation rate. The global
tendency
of decline in star formation rate with cosmological time is expected according to
simulations by Mayer et al.\ (2001). The values of $b$  by any of the morphological types
(excluding the interacting systems) spread roughly an order of magnitude, which probably 
indicates a great variety of star formation histories. The distribution of $b$ (see
Figure~\ref{histsfr} )is qualitatively similar to that found by van Zee  (2001).  We do
not notice any clear differences between the
different morphological types, but given the small sizes of the sample we have not tried
to quantify this accurately. 

Another way to compare the relevant time scales for the  star formation evolution is
by  computing how much time ($t_{\rm form}$) the galaxy would have needed to form all the
stellar
content at the current star rate, and what is the maximum time ($t_{\rm gas}$) that the
galaxy can  continue forming stars at the current rate. The analysis of
this in conjunction with optical luminosities and current star formation allow a first
approximation to the evolution of the galaxy over cosmic times. An accurate limit for
$t_{\rm gas}$ can be estimated  simply dividing the current gas content by the current
star formation rate. This of course would correspond to an ideal case in which all the
gas would be converted into stars, and no losses or accretion occurred. Using the  LEDA
database,\footnote{http://leda.univ-lyon1.fr/} we have compiled the existing
HI measurements  as a way to estimate the gas content in each galaxy. These are
presented in Table~3 column 4. The values of $t_{\rm gas}$ indicate that the current gas
content would be enough to fuel the star formation at the current star formation rate at
least during another Hubble time. For
about one third of the galaxies  $t_{\rm form}$  exceeds the Hubble time, so that we
conclude
again that on average the star formation rate was higher in the past. As expected, the
starbursting galaxies  NGC~2718b, NGC~4541e and NGC~5965a$_1$, and NGC~5965a$_2$ are among
the objects with the lowest formation times (also NGC~3154a, which is the satellite galaxy 
in
our sample closest to their parent). 

Previous studies (e.g., Casoli et al.\ 1996) have  shown that star formation rate in
spiral
galaxies is correlated with the mass of molecular and atomic gas. Our results confirm that
relation as  illustrated in Figure~\ref{sfrga}.  The correlation extends one and two 
orders of magnitude in gas mass and star formation rate respectively. NGC~2718b,  one of
the
interacting galaxies have the largest SFR in the sample and also  one of the galaxies
with higher gas content. This seems reinforce the scenario of significant accretion of gas
from NGC~2718a as outlined in previous section.

Figure~\ref{tiempos} presents $t_{\rm gas}$ vs.\ $t_{form}$, for our sample of satellite
galaxies
and compares these time scales with those for  other four samples taken from the
literature:
the Sculptor group dIrrs studied  by Skillman et al.\ (2003), the Local Group dIrrs of
Mateo
(1998), the gas-rich  low surface brightness galaxies studied by van Zee et al.\ (1997),
and  the isolated dIrrs of van Zee, (2000, 2001). The figure shows a clear direct
correlation between both time scales. The range of magnitudes and the
selection effects are different for each sample os that it is difficult to extract
statistical conclusiones, although all the samples tend to follow common tendencies. The
lower limit in $t_{\rm form}$ in our 
sample is $\sim$1 Gyr and

\section{Summary}
We have carried out a detailed analysis of narrow-band observations of the 
H$\alpha$ emission in a sample of 31 satellites  
orbiting giant spiral isolated galaxies. The objects studied span the  range 
$-19<M_B <-15$ mag. and were selected according to the relative 
brightness (at least 2.2 magnitudes fainter than their parents), projected distances
($<500$ kpc) and relative velocity ($<500$ km s$^{-1}$) from the primaries.
We have presented imaging and photometry in a narrow filter band covering the
position of the H$\alpha$ line. The results can be summarized as follows: 

$\bullet$ In all the spirals and irregular satellites (29 objects)  we detected $H\alpha$
fluxes above 1.15
10$^{-14}$ erg s$^{-1}$ cm$^{-2}$. The inferred current star formation rates are in
the range 0.006--3.66 M$_\sun$ yr$^{-1}$.

$\bullet$ There are three cases of clear interacting pairs. Four of the galaxies in
these pairs are among the objects with higher star-forming activity. In contrast, 
the only two galaxies of the sample that are not forming stars are also members of
these pairs. We do not detect H$\alpha$ emission in the filaments associated with
these interactions.

$\bullet$ The object with a largest current star formation rate 
(apart of the interacting systems) corresponds to the satellite galaxy NGC 3154a, 
which has the smallest projected distance from its progenitor (19 kpc) and the SFR
could be due to interaction with the parent galaxy.

$\bullet$ The median of the birth parameter $b$ is 0.68, indicating a fall in
activity with cosmic time, assuming perfect efficiency the current gas content of the
galaxies is enough to fuel this activity for more than another Hubble time.

\begin{center}
{\bf \Large Appendix A} 
\end{center}

Description of the H$\alpha$ emission maps of each galaxy:

$\bullet$ NGC~488c: This is a low surface brightness irregular galaxy in which we
notice four main knots with H$\alpha$ emission located in the outer parts of
the galaxy.

$\bullet$ NGC~772b:  What  seems a bright foreground star is 
projected through the E side of the galaxy.  The H$\alpha$ emission is concentrated in a
few discrete spots,   the two brightest being situated NE.

$\bullet$ NGC~772c: There are diffuse and discrete H$\alpha$ emission features. The
two brightest spots are situated roughly symmetrically with respect to the
galactic center.

$\bullet$ NGC~1517a: There are about ten H$\alpha$ regions, which are particularly
bright in the external parts of the galaxy. Two of the H$\alpha$ features through
the W are clearly differentiated from the main body of the disk.

$\bullet$ NGC~1620a: The galaxy is irregular with two--three plumes emerging from the
W. In one of these plumes there are a few discrete features with H$\alpha$ emission. 
The rest of the emission is concentrated in a few features distributed over the
full galaxy, 

$\bullet$ NGC~1961a: A few clumpy  small and faint structures are detected in the
image after continuum subtraction. 

$\bullet$ NGC~1961b: It is a luminous spiral satellite galaxy which has a bright
point-like structure close to the center. The H$\alpha$ continuum subtracted
image shows a diffuse H$\alpha$ emission  with a few over-imposed features 
concentrated along two chains
in direction NE--SW enclosing the  central part of the galaxy. The brightest spot
at  3.8 arcsec from the geometrical center was suspected to be a foreground
star but it is the dominant component in the continuum free image, so we think it
is a real galactic feature (or an intruder) experiencing a strong
starburst.

$\bullet$ NGC~1961c: This galaxy is a face-on spiral, which has a  very rich
structure in
H$\alpha$ emission with several extended features. The two more intense are
situated in the geometrical center and in a spot NW which is also obvious
in the $R$-band image. 

$\bullet$ NGC~2424b: This is a very irregular galaxy which shows a clumpy structure
in the broad-band image. The H$\alpha$ continuum free image shows about six
discrete spots, being the most intense the one located to the SW of the 
galactic center.

$\bullet$ NGC~2718a and NGC~2718b: These galaxies are a clear pair of interacting
satellites and have a sharp and  very straight bridge ($\approx$ $92^{''}$ = 25
kpc) connecting both galaxies. Close to NGC~2718a, and perpendicular to the main
axis there are what could be a small companion. The continuum-free image shows
that the only component with H$\alpha$ emission is NGC~2718b which has 3--4
intense knots.

$\bullet$ NGC~2775a: The two brightest H$\alpha$ emission features are projected
very close and are only partially resolved. The position  coincides with the
geometrical center of the galaxy. Most of the remaining emission is concentrated
in a few spots situated to the SE  of the galaxy.

$\bullet$ NGC~2775c: This galaxy shows an intense H$\alpha$ emission with a very
irregular spatial distribution concentrated mainly in the northern part. A
differentiated feature is to the SE of the galaxy. The broad-band image shows in
that position a plume emerging from the main body of the galaxy.

$\bullet$ NGC~2916a: The H$\alpha$ subtracted image shows an intense  and diffuse 
structure along the major axis. There are two less intense knots situated in
the E and W edges of the galaxy. 

$\bullet$ NGC~3043a: There is an extended diffuse structure where it is possible to
distinguish at least five differentiated regions distributed through  the full 
area (but avoiding the geometrical center) of the galaxy.

$\bullet$ NGC~3154a: The H$\alpha$ emission is spread over the full projected area of
the galaxy in the form of diffuse emission and discrete knots. The most intense
spot is located at the geometrical center. This is the galaxy in our sample
which has the smallest projected distance from its progenitor and (apart from the
interacting galaxies), the one with largest H$\alpha$ luminosity. Although this
could be related with a possible interaction with its parent, we
do not notice relevant signs of geometrical distortions in the satellite.

$\bullet$ NGC~3735a: The H$\alpha$ emission is concentrated in the E side of the
galaxy. In particular, there a bright feature at the geometrical center and a few
other less intense features in  structures resembling arms. In the broad-band image,
there is some evidence of a tail connecting this galaxy and another smaller one 
of unknown redshift situated to the NE. However, this last galaxy does not show
any H$\alpha$ emission.

$\bullet$ NGC~4030b: The majority of the H$\alpha$ emission is concentrated in four
discrete  spots. One is located at the center, and another is situated  NW
coincident with what we have identified as a possible small interacting galaxy. 
Unfortunately, the redshift of this last object is unknown.

$\bullet$ NGC~4541a: The broad-band image shows a spiral edge-on galaxy with two
small structures perpendicular to the major axis on both edges. The H$\alpha$
continuum-free image shows a diffuse structure that extends over most of
the  projected area of the galaxy, and some very clumpy structure. This emission
seems to be asymetrically distributed through the SW, where one of the features
mentioned in the broad-band image shows clear evidence of H$\alpha$ emission. We
think that this could be a small  interacting object that could be enhancing
the star formation in the neighboring regions of the main galaxy.

$\bullet$ NGC~4541b and NGC~4541e: This pair of galaxies are separated by $\sim$41
arcsec.  The galaxies have morphological types  E and S0/Sa respectively. In 
the broad-band  image there are two tidal tails emerging from NGC~4541b. One of
them is pointing directly to  NGC~4541e and it seems clear that its origin is the
interaction with this galaxy.  The other, pointing nearly in the opposite
direction, seems more extended and is possibly the relic of a previous passage of
this galaxy near NGC~4541e. The continuum-subtracted H$\alpha$ image  shows an
intense  structure in  NGC~4541e, while  NGC~4541b has no H$\alpha$ emission. 
The H$\alpha$ emission in NGC~4541e seems to be composed of at least four major
clumps located in the main body of the galaxy. A few faint features seem to
follow two arms in approximately opposite directions.  

$\bullet$ NGC~4725a: This is the largest satellite in our sample. The morphology
corresponds to a late spiral with clear signs of distortions and possibly dust
obscuration. In particular, a
plume emerges in the NE direction. The H$\alpha$ emission is concentrated along the
major axis with two bright spots, one approximately in the geometric center and
the other to the SW.

$\bullet$ NGC~5248a: The H$\alpha$ features are distributed over the full projected
area of the galaxy in about a dozen faint independent features.

$\bullet$ NGC~5248b: The brightest spot is located at the geometrical center with a
number of  diffuse structures distributed through the disk. 

$\bullet$ NGC~5899a: The galaxy shows a rich structure with diffuse H$\alpha$
features extending over the disk and three major discrete features, one in the
center and the other two in direction NW. A line of diffuse H$\alpha$ emission
seems to cross the galaxy in a direction perpendicular to the major
axis. We think this is an example of extra-planar diffuse ionized gas.   

$\bullet$ NGC~5962d: The H$\alpha$ emission features are distributed in 
numerous discrete features in the disk, avoiding the central part of the galaxy.

$\bullet$ NGC~5965a:  This corresponds really to two close galaxies as was noted by
Zwicky (1971) and Guti\'errez \& Azzaro (2004), separated by $\sim$7.4 arcsec. Following
the notation of that paper, we  denote them as NGC~5965a$_1$ and  NGC~5965a$_2$
(other authors have denoted the two members of this pair as  SBS~1533+574b and
SBS~1533+574a respectively). The continuum-subtracted image shows that both components
have H$\alpha$ emission. In the NW component it is  possible to  recognize at least
two irregular features. The isophotes of the NW component are elongated in the
direction of the companion. Additionally, the system is surrounded by a diffuse halo. 

$\bullet$ NGC~6181a: This is a late-type spiral with evident signs of distortion in
the broad-band image. The H$\alpha$ continuum free image shows a complex
structure dominated by a bright discrete feature which shows at least
three bright spots, It seems that this spot corresponds to a structure which
is differentiated from the main body of the galaxy and could correspond
to a minor merger.

$\bullet$ NGC~7137a: The H$\alpha$ emission is concentrated in two extended
features situated in the E side of the galaxy. The brightest runs in direction
perpendicular to the main axis.

$\bullet$ NGC~7678a: This nice face-on spiral galaxy has H$\alpha$ emission 
in the form of discrete features in the center and in a ring surrounding the
galaxy and approximately tracing  the spiral structure. A  small
galaxy is located very close by to the NW. This does not show any H$\alpha$
emission. The main H$\alpha$-emitting regions are located in two extended regions
situated in the NE part of the galaxy. Two plumes of diffuse material seem
connect these two structures with the main body of the galaxy. We think that they
could be two separate structures in the process of strong interaction with the
main galaxy.

\acknowledgments
We thank S. Laine who read the manuscript and given us very useful hints and comments. 
We thank also F. Prada who participated in the first stages of this project. 
E. D. Skillman has kindly provided the data used for the comparison presented in
Fig.\ref{tiempos}. S. A. was  supported by the Consejo Nacional de Investigaciones
Cient\'\i ficas y T\'ecnicas (CONICET), and by a Marie Curie Fellowship of the
European Community program MEST-CT-2004-504604.

\clearpage

\begin{deluxetable}{lcrccclr}
\tablewidth{0pt}
\tablecaption{Log of the observations}
\tablehead{
\colhead{Galaxy}           & \colhead{Epoch}      &
\colhead{t$_R$ }          & \colhead{t$_{H\alpha }$}  &
\colhead{FWHM R }    &  \colhead{FWHM ${H\alpha }$ }     & \colhead{Cal. star}    &
\colhead{H$\alpha$ Filter} \\
& & (s) & (s) & (arcsec) & (arcsec) }
\startdata 
NGC~488c     &  Dec 2001 &     1200          & 3x1800    & 2.2 & 1.8  & BD+75 325 &   663
\\
NGC~772b     &  Dec 2001 &     1200          & 3x1800    & 1.5 & 1.6  & BD+75 325 &   663
\\
NGC~772c     &  Dec 2001 &     1200          & 3x1800    & 1.3 & 1.7  & GD 50     &   663
\\ 
NGC~1517a    &  Dec 2001 &      900          & 3x1800    & 1.6 & 1.8  & BD+75 325 &   663
\\
NGC~1620a    &  Dec 2001 &      180          & 3x1800    & 1.6 & 2.1  & BD+75 325 &   663
\\
\\
NGC~1961a    &  Dec 2001 &      900          & 3x1800    & 1.6 & 1.6  & G193-74   &   663
\\
NGC~1961b    &  Dec 2001 &     1200          & 3x1800    & 1.7 & 1.4  & BD+75 325 &   663
\\
NGC~1961c    &  Dec 2001 &     1200          & 3x1800    & 1.1 & 1.2  & BD+75 325 &   663
\\
NGC~2424b    &  Dec 2001 &     1200          & 3x1800    & 1.6 & 1.6  &        ---    &  
663 \\
NGC~2718a    &  May 2002 &      900          & 3x1800    & 1.5 & 1.1  & Feige 34  &   663
\\
NGC~2718b    &  May 2002 &      900          & 3x1800    & 1.5 & 1.1  & Feige 34  &   663
\\
NGC~2775a    &  Dec 2001 &      300          & 3x1800    & 1.2 & 1.5  & BD+75 325 &   658
\\
NGC~2775c    &  Dec 2001 &     1200          & 3x1800    & 1.3 & 1.4  & BD+75 325 &   658
\\
NGC~2916a    &  Dec 2001 &     1200          & 3x1800    & 1.7 & 1.5  & BD+75 325 &   663
\\
NGC~3043a    &  Dec 2001 &     1200          & 3x1800    & 1.6 & 1.2  & BD+75 325 &   663
\\
NGC~3154a    &  Dec 2001 &     1200          & 3x1800    & 1.1 & 1.1  & BD+75 325 &   673
\\
NGC~3735a    &  Dec 2001 &     1200          & 3x1800    & 1.2 & 1.4  & BD+75 325 &   663
\\
\\
NGC~4030b   &  May 2002 &     1200          & 3x1800    & 1.4 & 1.7  & Feige 66  &   658
\\
NGC~4541a    &  May 2002 &     1200          & 3x1800    & 1.6 & 1.6  & Feige 34  &   673
\\
NGC~4541b    &  May 2002 &     1200          & 3x1800    & 1.6 & 1.6  & Feige 34  &   673
\\
NGC~4541e    &  May 2002 &     1200          & 3x1800    & 1.6 & 1.6  & Feige 34  &   673
\\
NGC~4725a    &  May 2002 &     1200          & 3x1800    & 1.2 & 1.3  & BD+33 2642&   658
\\
NGC~5248a    &  May 2002 &     1200          & 3x1800    & 1.1 & 1.2  & BD+33 2642&   658
\\
NGC~5248b    &  May 2002 &      600          & 3x1800    & 1.8 & 1.7  & Hz 44     &   658
\\
NGC~5899a    &  May 2002 &     1800          & 3x1800    & 1.4 & 1.4  & BD+33 2642&   663
\\
NGC~5962d    &  May 2002 &     1200          & 3x1800    & 1.1 & 1.3  & BD+33 2642&   663
\\
\\
NGC~5965a$_1$   &  May 2002 &      600       & 3x1800    & 1.4 & 1.2  & BD+33 2642& 663   
\\
NGC~5965a$_2$   &  May 2002 &      600       & 3x1800    & 1.4 & 1.2  & BD+33 2642& 663   
\\
NGC~6181a    &  May 2002 &      600          & 3x1800    & 1.2 & 1.2  & BD+33 2642&   663
\\
NGC~7137a    &  Dec 2002 &      500          & 3x1200    & 2.2 & 2.3  &        ---    &  
658  \\
NGC~7678a    &  Dec 2001 &     1200          & 3x1800    & 1.4 & 1.6  & BD+75 325 &   663
\\
\enddata
\end{deluxetable}

\clearpage

\begin{deluxetable}{lcccc}
\tablewidth{0pt}
\tablecaption{$H\alpha$ equivalent width, fluxes and  $H\alpha$ luminosities}
\tablehead{
\colhead{Galaxy} &
\colhead{$r_p$ }& \colhead{EW$_{H\alpha }$}& \colhead{H$_\alpha$ Flux}  &
\colhead{$\log$ L(H$\alpha$)}      \\
& (kpc)& (\AA) & ($10^{-14}$ ergs cm$^{-2}$ s$^{-1}$) & (ergs s$^{-1}$) }
\startdata  
NGC~488c     & 116&   8.07&  1.15   &    39.13          \\
NGC~772b     & 390&  13.97&  2.15   &    39.48          \\
NGC~772c     & 429&  15.19&  4.63   &    39.81          \\
NGC~1517a    & 130&  23.76&  9.33   &    40.41          \\
NGC~1620a    & 227&   8.03&  2.75   &    39.89          \\
NGC~1961a    & 214&   6.89&  7.97   &    40.45          \\
NGC~1961b    & 139&  15.01&  9.90   &    40.55          \\
NGC~1961c    & 120&  28.42& 29.00   &    41.01          \\
NGC~2424b    & 182&  35.52&         &                  \\
NGC~2718a    & 102& -11.64&         &                  \\
NGC~2718b    &  82& 118.58& 23.80   &    40.91          \\
NGC~2775a    & 401&  23.99&  9.53   &    39.61          \\
NGC~2775c    &  64&  19.26& 21.40   &    39.96          \\
NGC~2916a    &  71&  14.01&  2.66   &    39.93          \\
NGC~3043a    & 263&  38.72&  4.02   &    39.92          \\
NGC~3154a    &  19&  84.40& 13.00   &    41.11          \\
NGC~3735a    & 192&   3.54&  2.01   &    39.53          \\
NGC~4030b   & 414&  23.03&  1.46   &    38.85          \\
NGC~4541a    & 193&  22.89&  5.48   &    40.78          \\
NGC~4541b    & 228&  -1.29&         &                  \\
NGC~4541e    & 217& 101.22& 42.10   &    41.66          \\
NGC~4725a    & 132&  11.41& 49.80   &    40.22          \\
NGC~5248a    & 150&   9.05&  3.78   &    39.06          \\
NGC~5248b    & 164&  12.33&  2.76   &    38.93          \\
NGC~5899a    & 106&  19.14& 40.80   &    40.79          \\
NGC~5962d    &  89&   6.80&  3.58   &    39.50          \\
NGC~5965a$_1$& 532& 440.68& 16.50   &    40.64          \\
NGC~5965a$_2$& 532& 102.03&  9.41   &    40.40          \\
NGC~6181a    & 257&  30.49& 46.40   &    40.78          \\
NGC~7137a    &  64&  25.21&  2.24   &    39.17          \\
NGC~7678a    & 165&  18.15&  9.36   &    40.42          \\
\enddata
\tablenotetext{*}{Observation for NGC~2424a were not taken in photometric
conditions.}
\end{deluxetable}

\clearpage
\begin{deluxetable}{lcccrr}
\tablewidth{0pt}
\tablecaption{HI mass, star formation properties and time scales}
\tablehead{
 \colhead{Galaxy} &  \colhead{SFR}  & \colhead{$<$SFR$>_{Past}$} &\colhead{log  M$_{HI}$}
&
  
\colhead{$t_{gas}$}
&   \colhead{$t_{form}$} \\
& (M$_\sun$ yr$^{-1}$) &  (M$_\sun$ yr$^{-1}$) & (M$_\sun$ ) &  (Gyr) & (Gyr)}
\startdata 
NGC~488c     &  0.011   &    0.048  & 8.75       &     51.9&     12.9 \\
NGC~772b     &  0.024   &    0.044  & 8.78       &     25.1&      5.8 \\
NGC~772c     &  0.051   &    0.107  & 8.80       &     12.4&      5.1 \\
NGC~1517a    &  0.206   &    0.290  & 9.57       &     17.9&      2.7 \\
NGC~1620a    &  0.062   &    0.149  & 9.35       &     35.5&     10.7 \\
NGC~1961a    &  0.225   &    0.867  & 9.73       &     23.5&     12.9 \\
NGC~1961b    &  0.279   &    0.447  &        &       &        5.5 \\
NGC~1961c    &  0.817   &    0.938  &        &       &        3.6 \\
NGC~2424b    &          &    0.048  & 9.05       &       &            \\
NGC~2718a    &          &    0.221  &        &       &            \\
NGC~2718b    &  0.640   &    0.081  & 9.50       &      4.8&      1.8 \\
NGC~2775a    &  0.032   &    0.050  & 9.06       &     37.6&     13.1 \\
NGC~2775c    &  0.072   &    0.123  & 8.74       &      7.8&      8.4 \\
NGC~2916a    &  0.067   &    0.285  & 9.34       &     32.5&     17.2 \\
NGC~3043a    &  0.066   &    0.073  &        &       &            \\
NGC~3154a    &  1.016   &    0.347  &        &       &        1.5 \\
NGC~3735a    &  0.027   &    0.222  &        &       &       36.1 \\
NGC~4030b    &  0.006   &    0.132  & 8.74       &     95.0&    128.7 \\
NGC~4541a    &  0.476   &    0.486  &        &       &       7.3 \\
NGC~4541b    &          &    0.382  &        &       &        \\
NGC~4541e    &  3.659   &    1.043  &        &       &      1.0 \\
NGC~4725a    &  0.132   &    0.336  & 9.21       &     12.0&     12.6 \\
NGC~5248a    &  0.009   &    0.041  & 8.82       &     70.6&     19.9 \\
NGC~5248b    &  0.007   &    0.057  & 8.45       &     41.7&     27.3 \\
NGC~5899a    &  0.489   &    0.434  & 9.56       &      7.3&      4.5 \\
NGC~5962d    &  0.025   &    0.230  & 8.56       &     14.4&     35.1 \\
NGC~5965a$_1$&  0.350   &    0.132  &        &         &      1.7  \\
NGC~5965a$_2$&  0.200   &    0.058  &        &         &      1.3  \\
NGC~6181a    &  0.476   &    0.421  & 9.36       &      4.8&      3.8 \\
NGC~7137a    &  0.012   &    0.024  & 8.80       &     54.0&      7.5 \\ 
NGC~7678a    &  0.208   &    0.400  & 9.39       &     11.5&      6.7\\
\enddata
\tablenotetext{*}{HI data were collected from the LEDA catalogue}
\end{deluxetable}

\clearpage
\onecolumn

\begin{figure}
\plotone{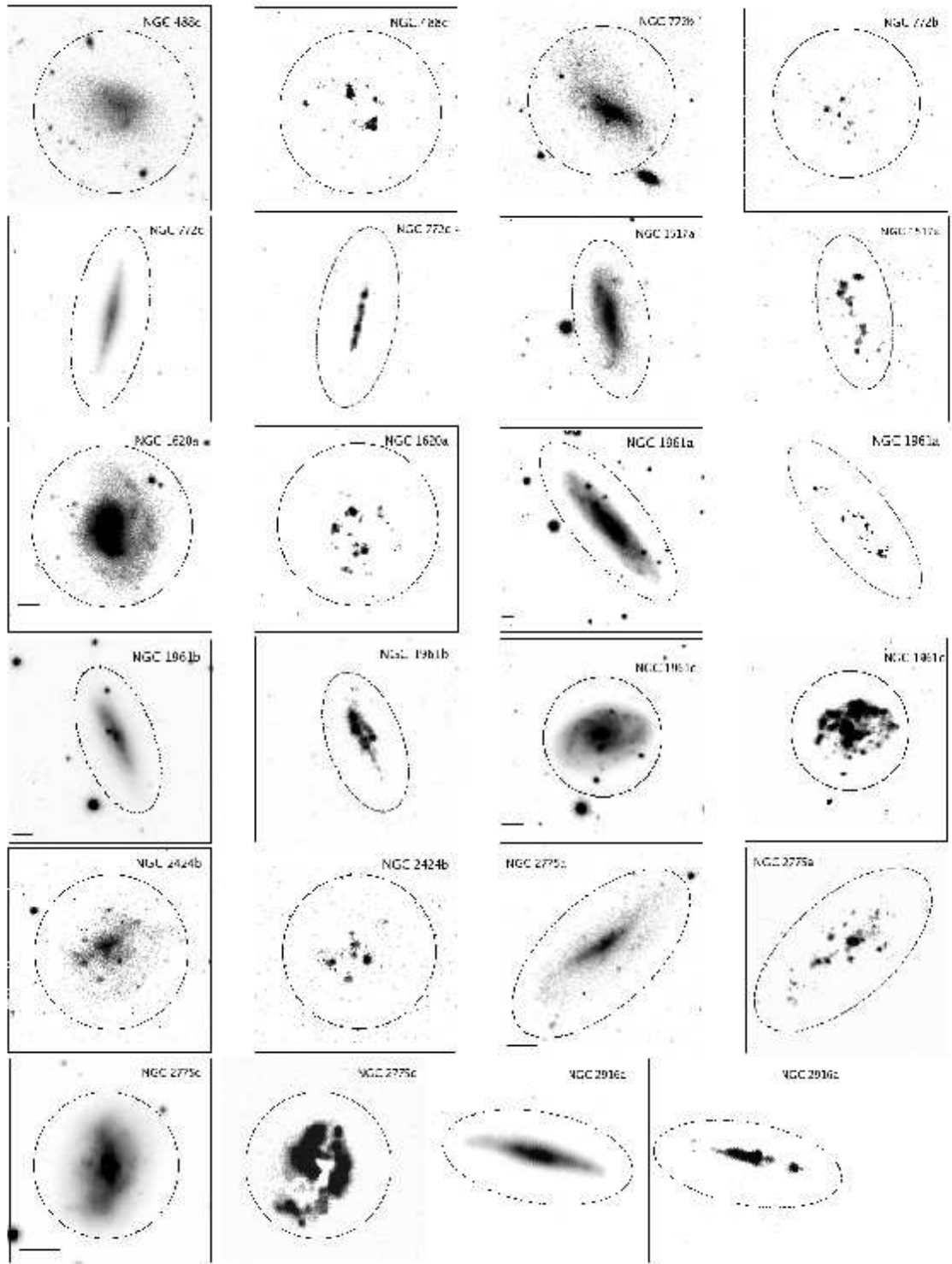}
\caption{Images of the sample of objects analysed in this work. For each object are
presented the R-band  and H$\alpha$ images after continuum subtraction.
North is up and east to the left. 
The small line at the bottom is scaled to an angular size of 10 arcsecs.
}
\end{figure}

\begin{figure}
\plotone{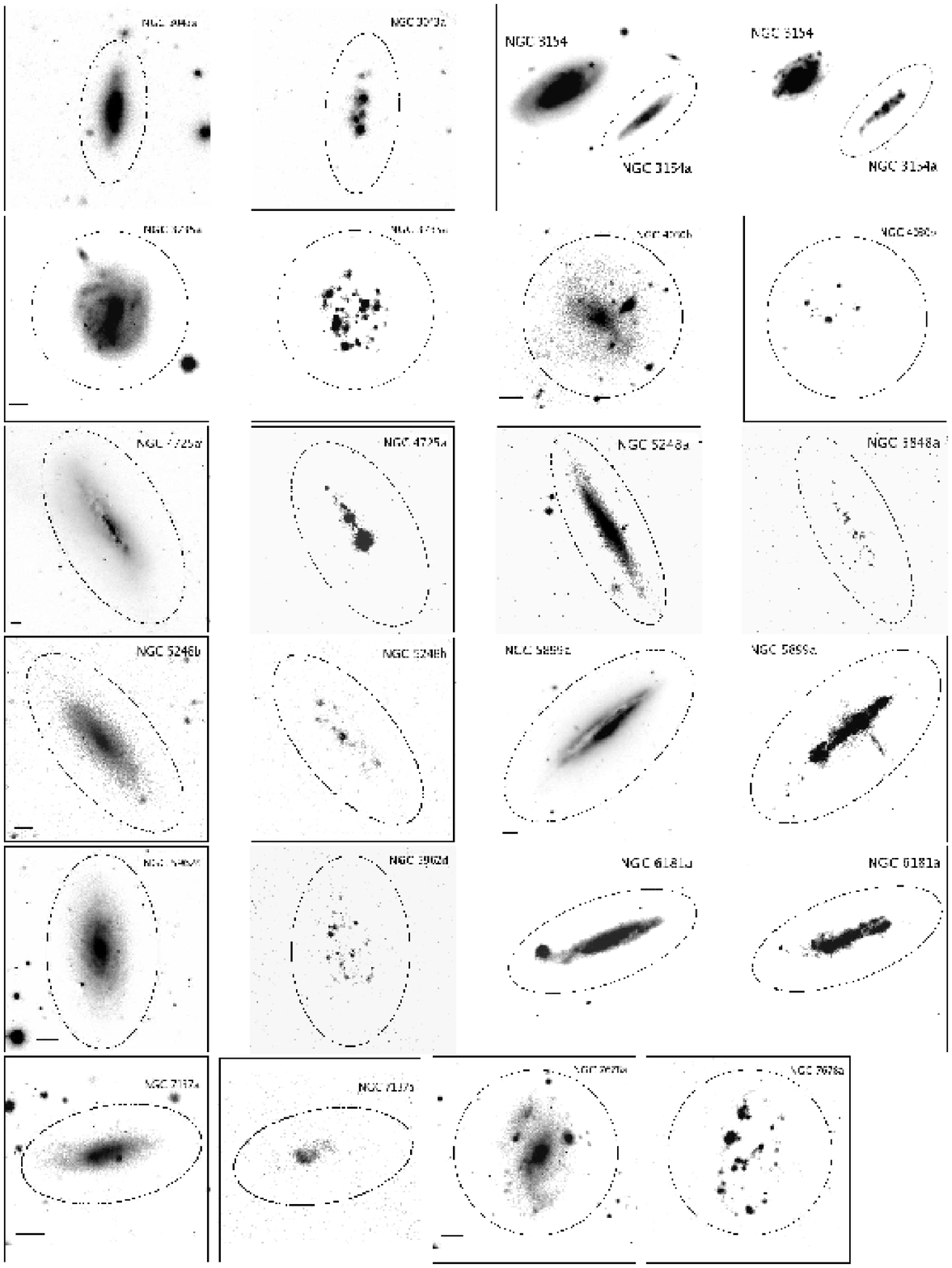}
\caption{Cont.}
\end{figure}

\begin{figure}
\plotone{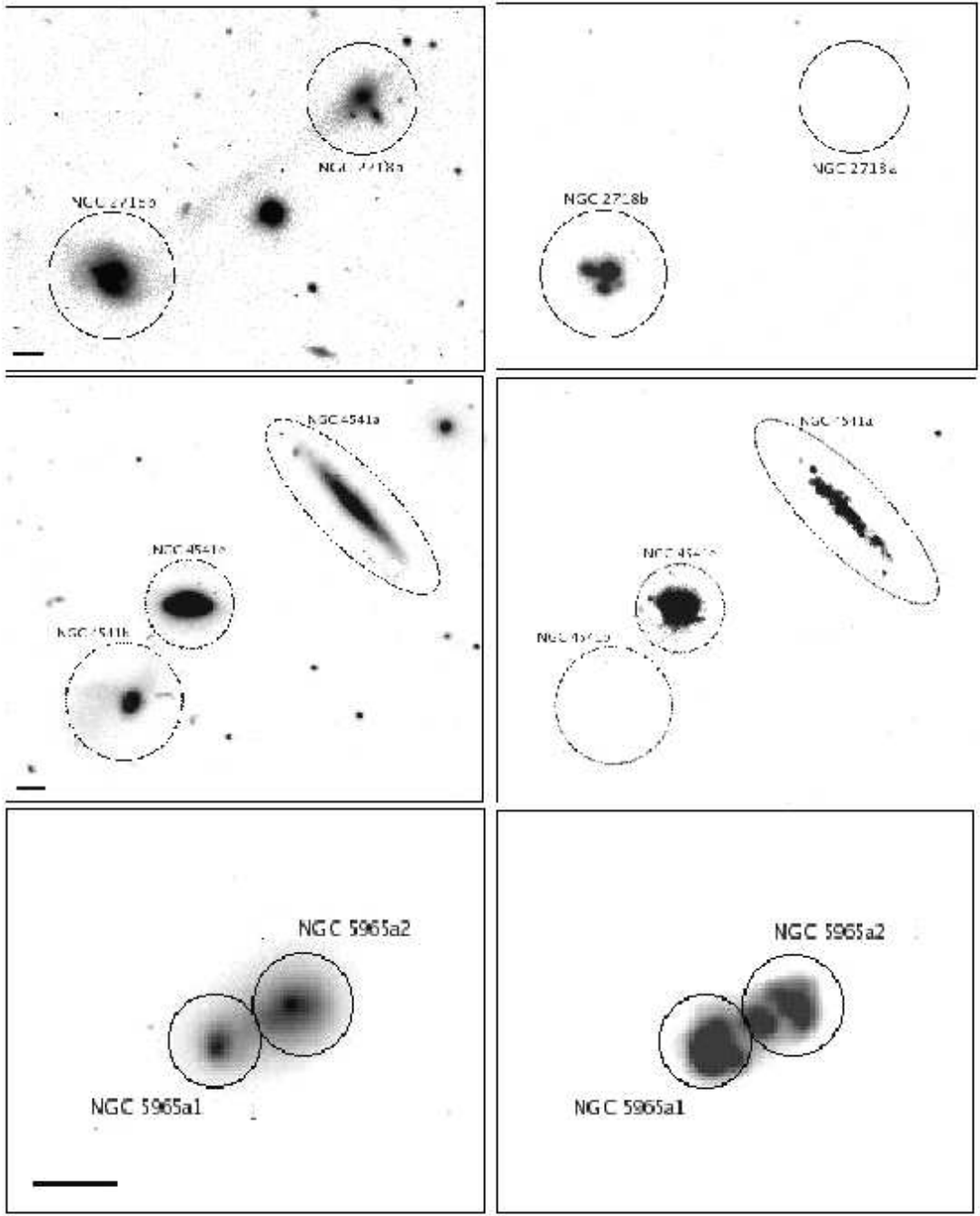}
\caption{Idem for the three pairs of interacting satellite galaxies: NGC2718a-b,
NGC4541a, NGC4541b-e 
and NGC5965a$_1$-a$_2$.}
\end{figure}

\begin{figure}
\plotone{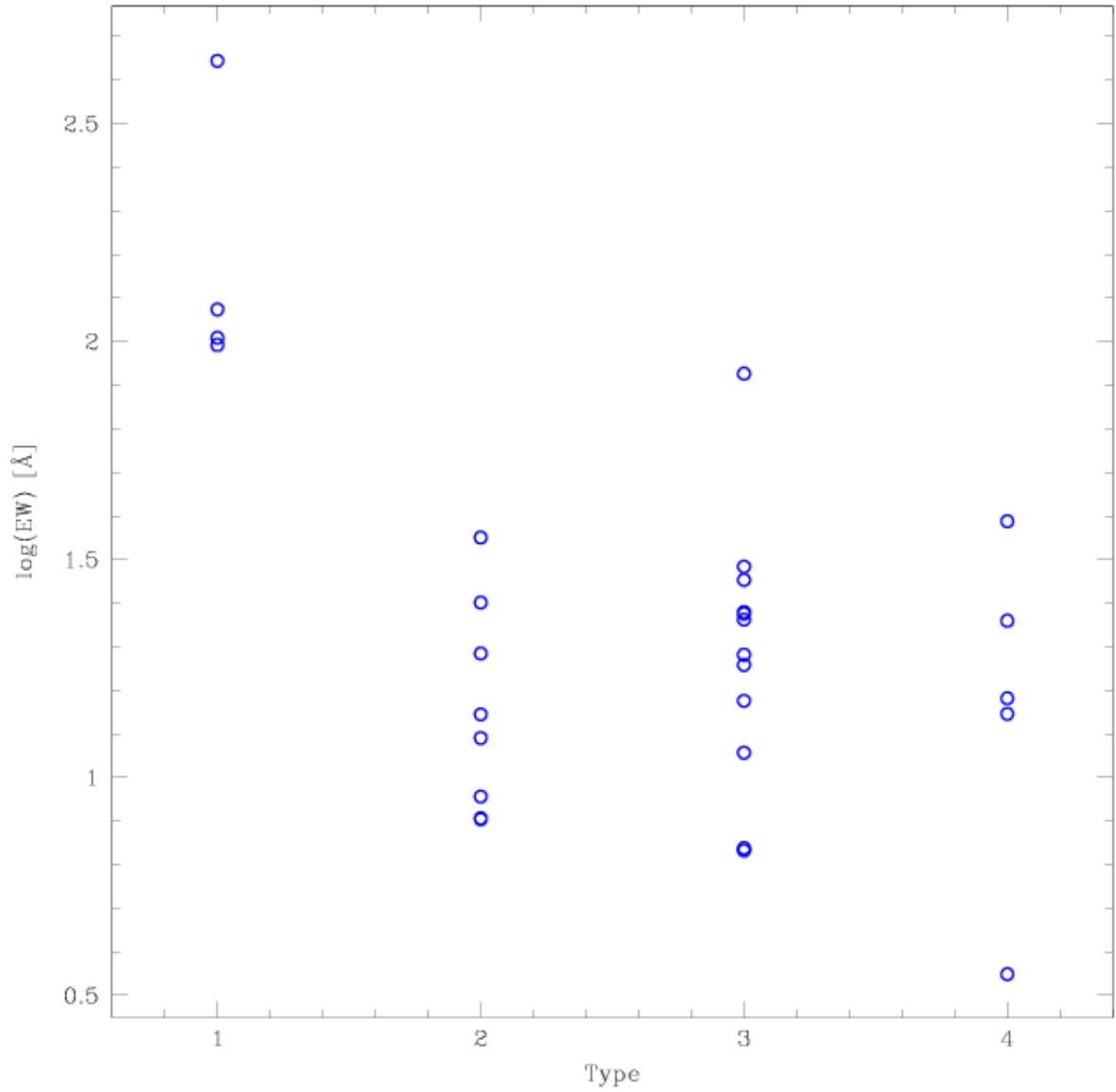}
\caption{$H\alpha$ equivalent widths as a function of the galactic type: 
(1) Objects in interaction, (2) Irregular, (3) Sb/Sc and (4) Sa. 
}
\label{ewtipo}
\end{figure}

\begin{figure}
\plotone{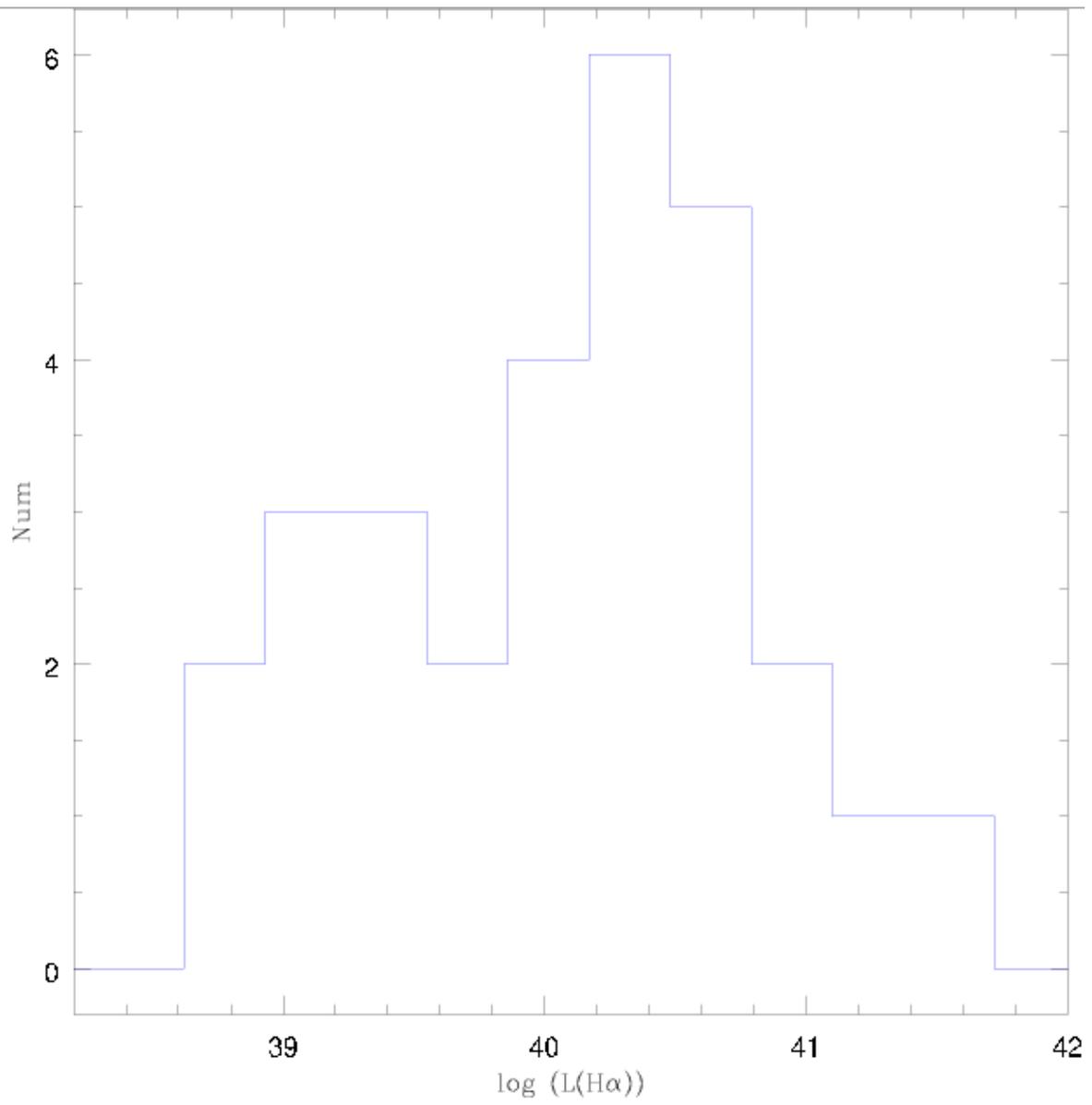}
\caption{Distribution of $H\alpha$ luminosities for the sample of satellite galaxies
analysed in this paper.}
\label{lumha}
\end{figure}

\begin{figure}
\plotone{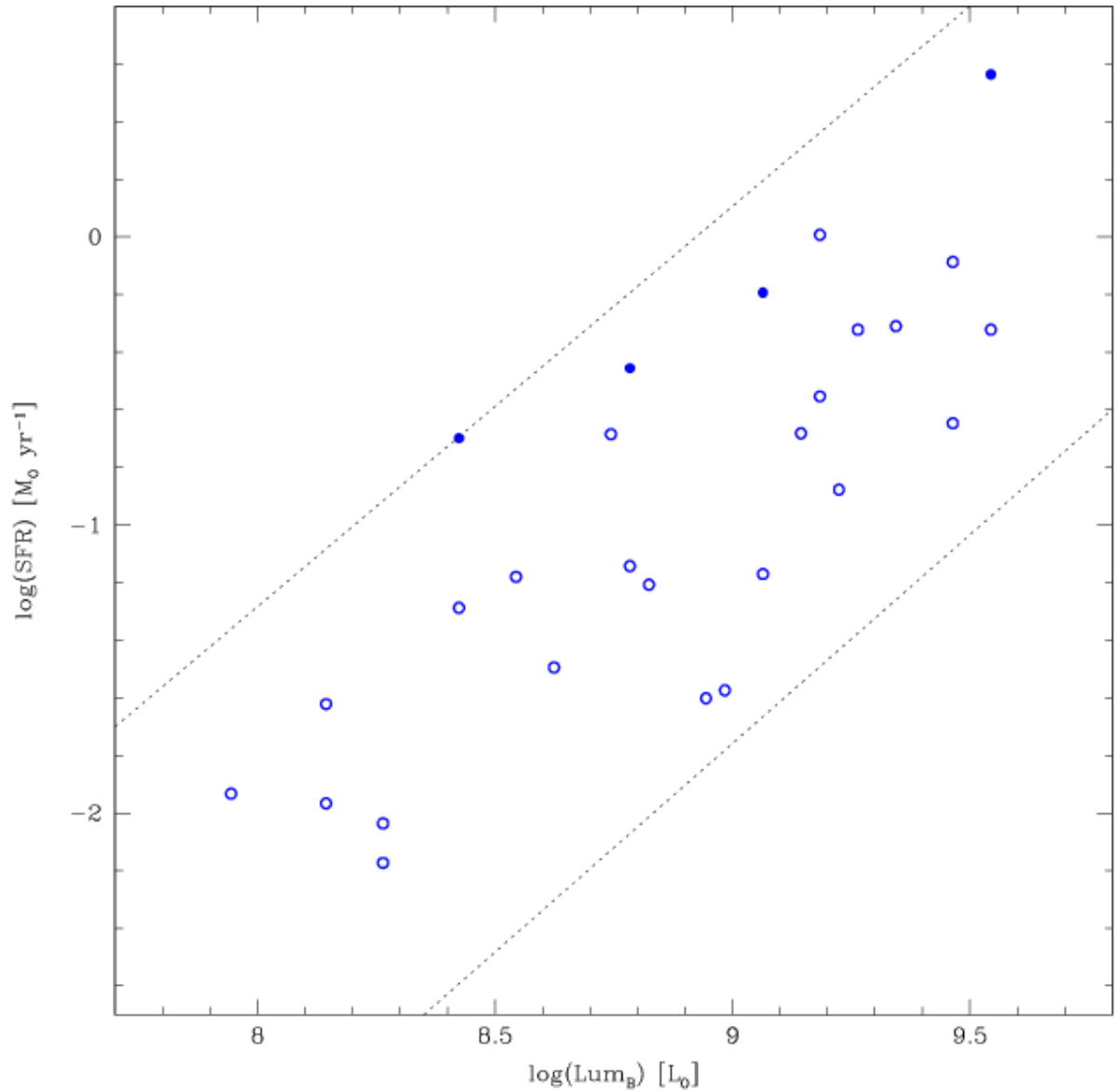}
\caption{Current star formation rate, $SFR$, estimated from the $H\alpha$ luminosities, 
as a function of luminosity in the $B-band$.
Objects in interaction have been indicated by a filled circle. 
}
\label{sfrb}
\end{figure}

\begin{figure}
\plotone{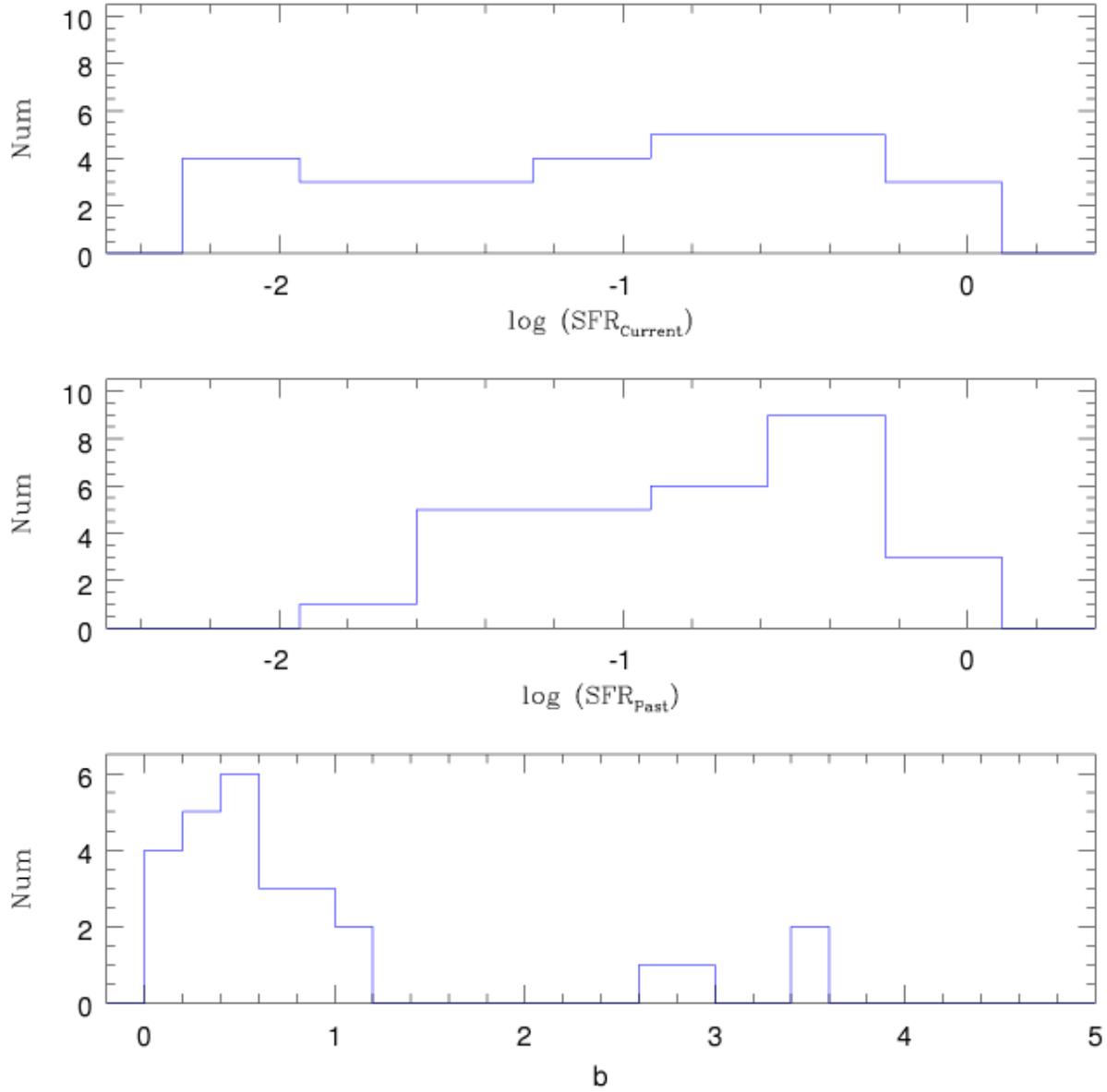}
\caption{Distribution of current ($top$) and past ($medium$) star formation rates,
for the sample of satellite galaxies analysed in this paper.
The bottom box shows the distribution of stellar birthrate parameter, 
$b$ $=$ $SFR_{Current}/<SFR>_{Past}$ .  
}
\label{histsfr}
\end{figure}

\begin{figure}
\plotone{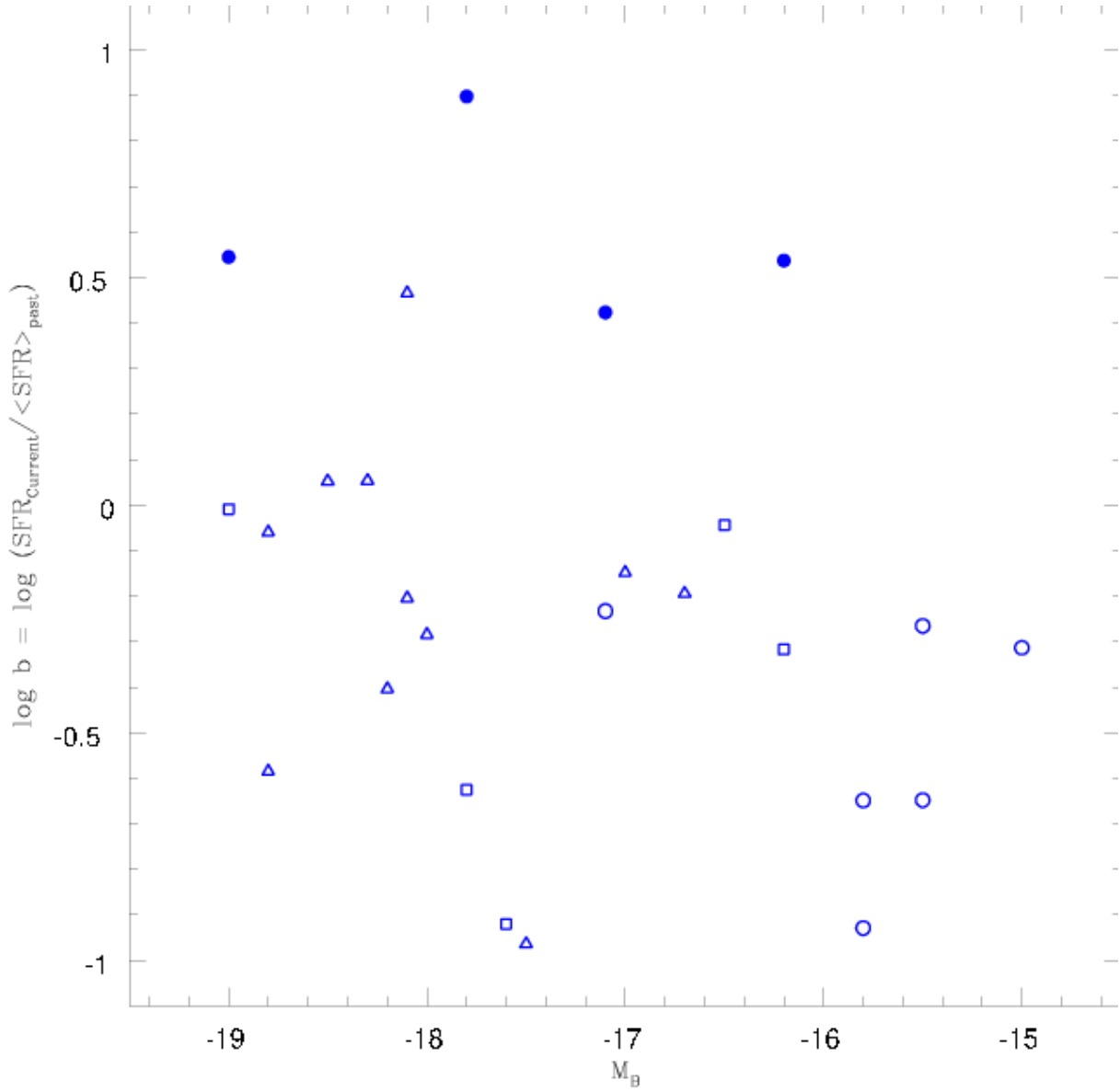}
\caption{Stellar birthrate parameter as a function of the absolute 
magnitude ($M_B$), in satellite galaxies for different Hubble morphological types: 
Irregulars ({\it open circles}), Sb/Sc ({\it open triangles}) and  Sa ({\it open
squares}).
Interacting galaxies are indicated by filled symbols.}
\label{bscalo}

\end{figure}
\begin{figure}
\plotone{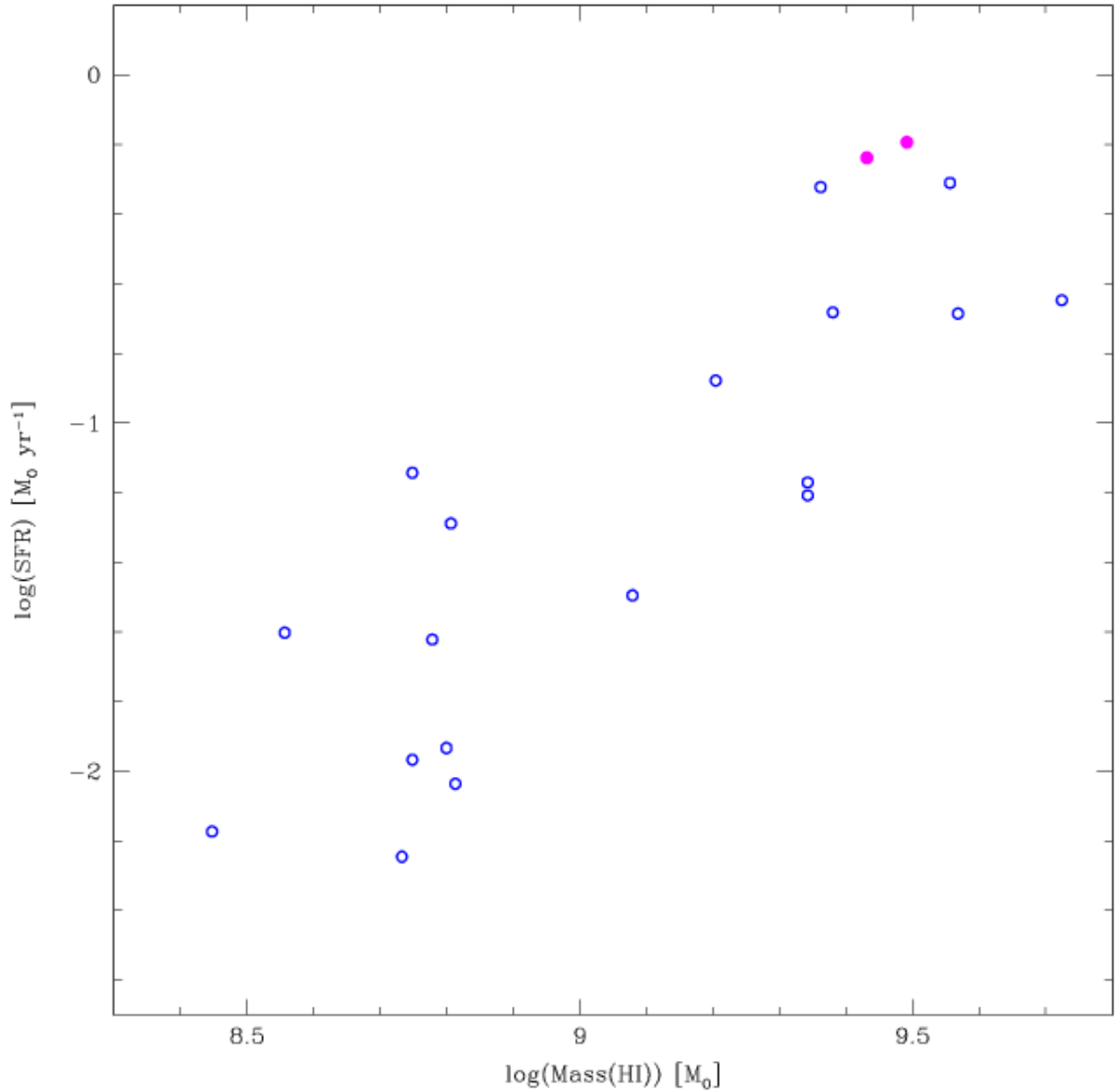}
\caption{Current star formation rate, $SFR$, estimated from H$\alpha$ measurement, vs. 
HI mass of the gas. Interacting galaxies, NGC 2718b and NGC 5965a (computed in an aperture
enclosing both galaxies, NGC 5965a$_1$ and a$_2$), are indicated by filled symbol.}
\label{sfrga}
\end{figure}

\begin{figure}
\plotone{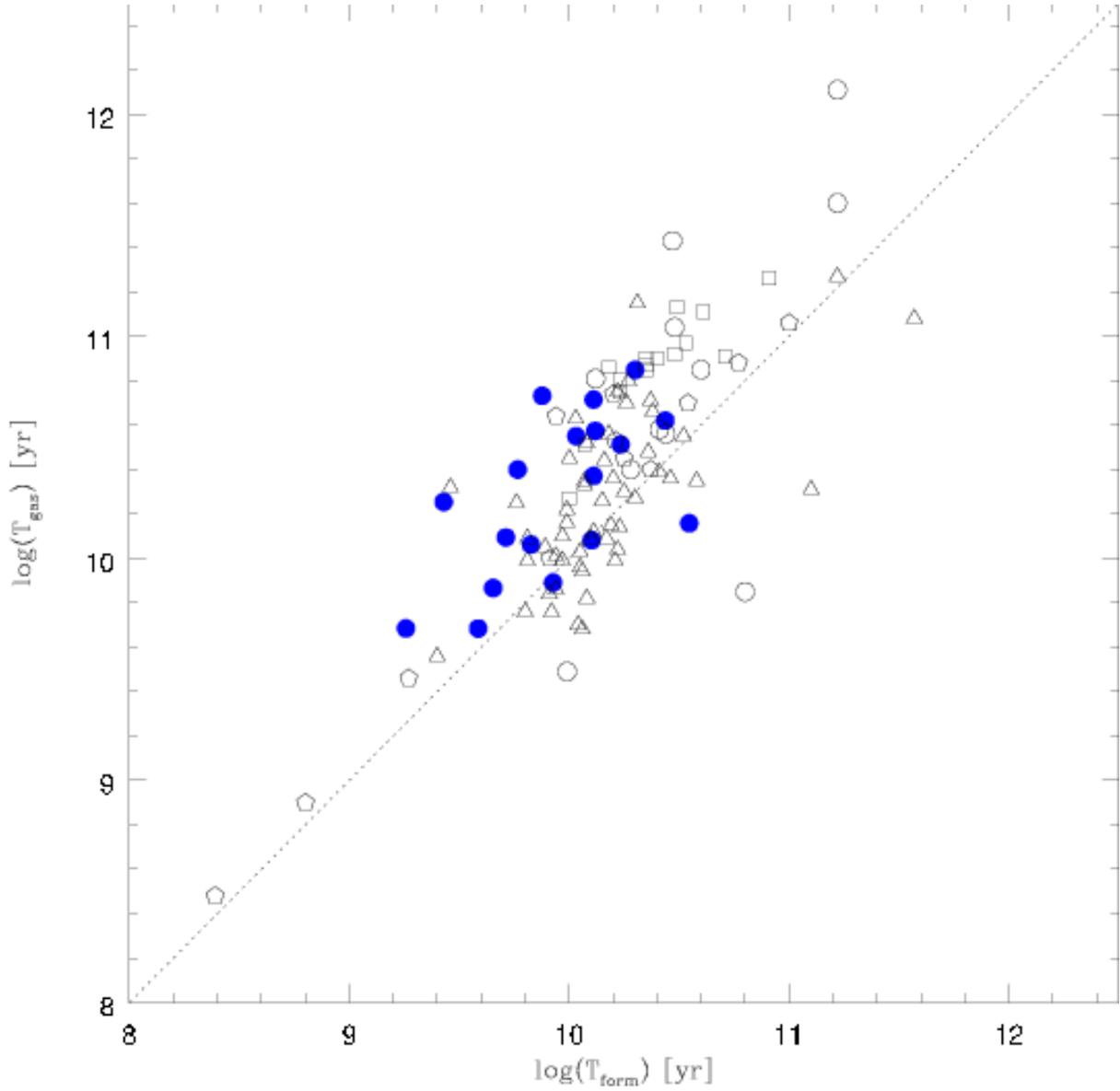}
\caption{Comparison between the ratio of the gas mass to the current $SFR$ ($T_{gas}$)
and the ratio of the luminosity to the current $SFR$ ($T_{form}$), for the satellite
galaxies (filled circles), and four samples taken from the literature: the Sculptor 
group dIrrs studied by Skillman et al. (2003) (open circles), the Local Group dIrrs of
Mateo (1998) (open pentagons), the gas rich low surface brightness galaxies studied
by van Zee et al. (1997) (open squares), and the isolated dIrrs of van Zee, 
(2000, 2001) (open triangles).
}
\label{tiempos}
\end{figure}

\end{document}